\documentclass[aps,prl,twocolumn,superscriptaddress,floatfix]{revtex4-1}
\usepackage{amsmath,amssymb,graphicx,graphics,color}
\usepackage{dcolumn}
\usepackage{bm}
\usepackage{psfrag}

\begin{document}

\title{ Ground-state phases of rung-alternated spin-1/2 Heisenberg ladder}
\author{F. Amiri}
\affiliation{Department of Physics, University of Guilan, Rasht, Iran}
\author{G. Sun}
\affiliation{Max Plank Institut f\"ur Physik komplexer Systeme, Dresden, Germany}
\author{H.-J. Mikeska}
\affiliation{Institut f\"ur Theoretische Physik, Leibniz Universit\"at Hannover, Germany}

 \author {T. Vekua}
\affiliation{Institut f\"ur Theoretische Physik, Leibniz Universit\"at Hannover, Germany}

\begin{abstract}

 The ground-state phase diagram of Heisenberg spin-1/2 system on a two-leg ladder with rung alternation is studied by combining analytical approaches with numerical simulations. For the case of ferromagnetic 
leg exchanges a unique ferrimagnetic ground state emerges, whereas for the case of antiferromagnetic leg exchanges several different ground states are stabilized depending on the ratio between exchanges along legs and rungs. For the more general case of a honeycomb-ladder model for the case of ferromagnetic leg exchanges besides usual rung-singlet and saturated ferromagnetic states we obtain a ferrimagnetic Luttinger liquid phase with both linear and quadratic low energy dispersions and ground state magnetization continuously changing with system parameters. For the case of antiferromagnetic exchanges along legs, different dimerized states including states with additional topological order are suggested to be realized. 
\end{abstract}

\maketitle

\date{\today}

% \pacs{05.30.Fk, 03.75.Ss, 03.75.Mn, 71.10.Fd}

\section{Introduction}

Spin-1/2 Heisenberg two-leg ladder systems have attracted a great deal of interest both from experiment and from theory\cite{KM}. 
Ladders with antiferromagnetic exchange along rungs and antiferromagnetic \cite{Takano91,TMR93,Eccleston94} as well as ferromagnetic \cite{Yamaguchi,Yama2014} exchanges along legs have been realized experimentally. 

For the case of a ladder with ferromagnetic legs and a uniform inter-leg (rung) exchange, so called rung-singlet or saturated ferromagnetic phases are realized depending whether the inter-leg coupling is antiferromagnetic or ferromagnetic.

For the case of a ladder with antiferromagnetic legs and a uniform rung coupling it has been established that for antiferromagnetic inter-leg coupling a rung-singlet phase is realized, whereas for ferromagnetic rung coupling a Haldane phase is stabilized. Both phases are stable both in weak rung-coupling and in strong rung-coupling limits. Weak rung-coupling limit is a proper limit for effective field theory bosonization analyses \cite{GNT}, where in the case of uniform rung exchanges, in the lowest (first) order of the inter-chain coupling the relevant operators (in the renormalization group sense) are present that at low energies drive system towards the strong-coupling fixed points of the rung-singlet and Haldane states for positive and negative rung exchanges respectively. 

Stability of a unique ground state from arbitrary weak (non-zero) up to arbitrary strong inter-leg exchanges, is due to the fact that the ladder system is non-frustrated. The question that we are going to address in our work is what happens when the ladder system is frustrated by rung exchange alternating in sign from rung to rung. Frustration in this case, for both signs of exchanges along the ladder legs, will be caused by the presence of an odd number of antiferromagnetic exchanges in the elementary closed path that is a ladder plaquette in our case. This problem has not been addressed before and we will try to fill this gap in the following.

We will use different complementary analytical approaches: strong-rung coupling expansion for strongly coupled legs and bosonization for weakly coupled antiferromagnetic legs. To cover the intermediate regimes we will use numerical techniques. We will as well consider the more generalized case of a rung-alternated model where we will relax the constraint of equal absolute value of exchanges along the even and odd rungs.

\section{Model of frustrated spin ladder}

\begin{figure}%[ht]
\includegraphics[width=7.50cm]{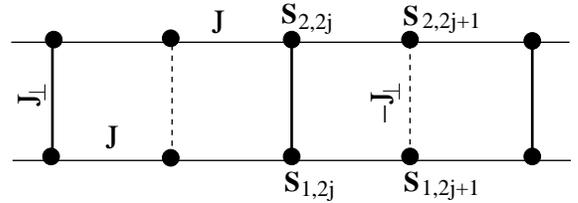}
\caption{ Geometry of the two-leg ladder with alternated rung exchanges. Antiferromagnetic couplings along are chosen along the even rungs, $J_{\bot}>0$.  }
\label{fig:model1}
\end{figure}

In this work we study a Heisenberg spin-1/2 model defined on a two leg ladder with $L$ rungs and with alternating rung exchanges, depicted in Fig. \ref{fig:model1},
%%%%%%%%%%%%%
\begin{eqnarray}
\label{Themodel}
H=  J\sum^L_{j=1,l}  {\bf S}_{l,j}{\bf S}_{l,j+1}+ J_{\bot}\sum^L_{j=1} (-1)^j{\bf S}_{1,j}{\bf S}_{2,j}\,\,,
\end{eqnarray}
%%%%%%%%%%%%%
where  ${\bf S}_{l,j}$ are spin-$\frac{1}{2}$ operators acting on spins on the $j$-th rung of the $l=1,2$ leg. For definiteness we will put $J_{\bot}\ge 0$, the case of $J_{\bot}\le 0$ will be recovered by one-site translation of the ladder along the legs.

The anisotropic $XY$ case of Eq.~(\ref{Themodel}) was studied recently in the context of single-component hard-core bosons on {a} two-leg ladder at half-filling with a flux $\pi$ per plaquette~\cite{Pir,Orignac}. It was shown that depending on the ratio of strengths of exchanges along ladder legs and rungs there are two different ground-state phases. For $|J|\ge J^c_{xy}$, where $J^c_{xy}\simeq 2/3 J_{\bot}$ and for the $XY$ case the sign of $J$ is irrelevant {\and} the ground state was shown to be a  vortex-liquid Mott insulator, a state with gapped magnetic excitation (excitations changing total $S^z$), however having a gapless mode to non-magnetic excitation. For $|J|\le J^c_{xy}$  the ground state was shown to be a fully gapped non-degenerate state, adiabatically connected to the $J=0$ case where the ground state is of simple product form and is composed of alternating $S^z=0$ components of triplets and singlets from rung to rung.
 {The presence} of multiple ground states when changing a single parameter (here ratio of exchanges along legs and rungs) is an indicator of frustration present in the system.

\section{Strong rung-coupling limit}

Let us start from the limit $|J|\ll J_{\bot}$. For $J=0$, even rungs in the ground state form rung-singlet states, whereas odd rungs form rung-triplet states. Hence for $J=0$ the ground state manifold is 
$3^{L/2}$ times degenerate. Treating $J$ perturbatively, integrating out singlets that occupy even rungs, we obtain an effective Hamiltonian describing a collection of odd rungs,
\begin{equation}
\label{Spin1}
H^{1}_{eff}= \frac{J^3}{16J^2_{\bot}}\sum^{L/2}_{j=1} {\bf T}_j{\bf T}_{j+1}+O(J^4) \,,
\end{equation}
where ${\bf T}_j= {\bf S}_{1,2j+1}+{\bf S}_{2,2j+1}$ are effective $S=1$ spins formed along the odd rungs. Because the exchanges along both legs have equal strengths the lowest, second order in $J$ contribution to the effective spin-1 chain formed on odd rungs vanishes. The ground state of our model Eq.(\ref{Themodel}) for $|J|\ll J_{\bot}$ will be hence a direct product of singlets formed on even rungs and the ground state of a Heisenberg spin-1 chain formed on odd rungs. Depending on the sign of $J$ the ground state of this Heisenberg spin-1 chain is either ferromagnetic state ($J< 0$), or Haldane state \cite{Haldane83} ($J> 0$). For the original Hamiltonian Eq.(\ref{Themodel}) ferromagnetic state of the effective $S=1$ chain Eq.(\ref{Spin1}) is the half-ferromagnetic state with the ground state total spin equal to half of the maximum possible value, $S^T=L/2$. This state will be called half-Ferro state. The effect of anisotropies on the Haldane state realized for $0<J\ll J_{\bot}$ has been studied recently \cite{Tonegawa}. For the rung-alternated ladder with $J>0$ it was shown that the application of an external magnetic field induces the half-magnetization plateau state \cite{Japa2006}.

\section{Weak rung-coupling limit, Bosonization}
In the other limit $J\gg J_{\bot}$ we use bosonization approach\cite{GNT}. We represent spin operators with the help of bosonic operators:
\begin{equation}
{\bf S}_{l,j} \to a[ {\bf J}_{l,L}(x)+ {\bf J}_{l,R}(x)+(-1)^{j} {\bf N}_l(x)].
\end{equation}
Decoupled chains have N\'eel-like quasi long-range order and the above
representation captures the important low energy fluctuations by smooth bosonic fields, at wave-vector $0$ and $\pi$.
Uniform spin magnetization is represented in terms of chiral currents ${\bf
  J}_{l,L/R}$ of {\bf a} level-$1$ $SU(2)$ Wess-Zumino-Witten model perturbed
by marginally irrelevant current-current interactions, describing an isolated antiferromagnetic Heisenberg chain\cite{Affleck85}.

We will need the following important operator product expansion (OPE) rules\cite{Balents},
\begin{equation}
\label{OPE}
J^a_{l,L/R}(x,\tau)N_l^b(x',\tau')=\frac{\pm i \delta_{ab}\epsilon_l (x',\tau')+i\epsilon_{abc}N_l^c(x',\tau') }{  4\pi [v(\tau-\tau') \pm i(x-x')]},
\end{equation}

where $v=\pi J/2$ is the spin-wave velocity of {  the} Heisenberg spin-1/2
chain, known from {  the} Bethe ansatz solution, and {  on} the right hand side {  the} dimerization operator $\epsilon_l$, that is the continuum limit of 
$(-1)^j{\bf S}_{l,j}{\bf S}_{l,j+1}$ has appeared.

Treating the inter-chain coupling $J_{\bot}$ perturbatively for $J_{\bot}/ J\ll 1$ in the continuum limit the staggered inter-chain coupling, $ H_{\bot}=J_{\bot}\sum^L_{j=1} (-1)^j{\bf S}_{1,j}{\bf S}_{2,j}$, has the following form in terms of smooth bosonic fields,
\begin{eqnarray}
H_{\bot}= && \int \mathrm{d}x \mathcal{H}_{\bot} (x)={J_{\bot}} \int \mathrm{d}x  ({\bf J}_{1,L}(x)+ {\bf J}_{1,R}(x))  {\bf N}_2(x). \nonumber \\
&&+{J_{\bot}} \int \mathrm{d}x ({\bf J}_{2,L}(x)+ {\bf J}_{2,R}(x))  {\bf N}_1(x) .
\end{eqnarray}
The $H_{\bot}$ perturbation has non-zero conformal spin and does not open a gap in the first order of $J_{\bot}$. The relevant scalar operator from inter-chain exchange comes in the second order of $J_{\bot}$ coupling.

Using OPE for the same-leg operators at short-distances Eq.(\ref{OPE}) and integrating with the relative coordinates we obtain in the order $J^2_{\bot}$ the following relevant contributions that should be added to effective Hamiltonian describing the long wavelength properties of decoupled bosonic chains,
\begin{equation}
\label{relevant}
\sim- J^2_{\bot}\int \mathrm{d}x [ 3\epsilon_1(x)\epsilon_2(x)  -2{\bf N}_1(x){\bf N}_2(x) ].
\end{equation}
At this point it is convenient to introduce 4 Majorana fermions \cite{Shelton}. The perturbation that we identified in Eq.(\ref{relevant}) is translated as mass term of triplet and singlet Majorana fermions and the complete ladder Hamiltonian in Majorana basis {  looks},
\begin{eqnarray}
H_M&&= \int  \mathrm{d}x [ \sum^3_{\gamma=1}  \frac{iv_t}{2} ( \psi_L^{\gamma} \partial_x \psi_L^{\gamma}-\psi_R^{\gamma} \partial_x \psi_R^{\gamma}) +  im_t \psi_L^{\gamma}\psi_R^{\gamma}]\nonumber \\
&&+   \int  \mathrm{d}x [ \frac{iv_s}{2} ( \psi_L^{0} \partial_x \psi_L^{0}-\psi_R^{0} \partial_x \psi_R^{0})   +  im_s \psi_L^0 \psi_R^0],
\end{eqnarray}
where the masses of Majoranas are: $m_t\sim 5J^2_{\bot}$ and $m_s\sim 3J^2_{\bot}$. Following the analyses of the ground states of the original ladder system from Majoranas \cite{GNT,Balents} we can conclude that the ladder system for $J_{\bot }/J\to 0$ is in the phase that is adiabatically connected to the rung-singlet phase of the uniform ladder with antiferromagnetic exchanges. This is surprising, because the spin exchanges are ferromagnetic along every other rung.

\begin{figure}%[ht]
\includegraphics[width=8.0cm]{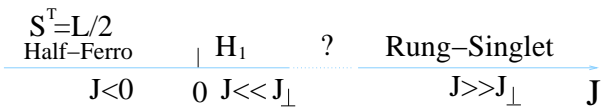}
\caption{ Analytically conjectured ground-state phase diagram of
  rung-alternated Heisenberg ladder as function of $J$. H$_1$ denotes a
  Haldane phase, however only half of the rungs (odd numbered rungs) provide the effective $S=1$ spins, whereas the other half (even numbered rungs) are in the approximate rung-singlet states and mediate antiferromagnetic exchange among effective $S=1$ spins.}
\label{fig:modelest}
\end{figure}

Comparing the effective Hamiltonian Eq. (\ref{Spin1}) and the original one Eq. (\ref{Themodel}) reveals a very important competition for$J>0$. Namely, for small $0<J \ll J_{\bot}$ the next nearest neighbour spins on the same chain belonging to the odd rungs show antiferomagnatictendencies, $\langle {\bf S}_{l,2j+1}{\bf S}_{l,2j+3} \rangle <0$, whereas
for the single-chain dominated regime $J\gg J_{\bot}$, it is clear that
$\langle {\bf S}_{l,j}{\bf S}_{l,j+2} \rangle >0$ for any $l$ and $j$. There is no such competition on $J<0$ side and hence the case of
ferromagnetic legs is simpler and for the entire region of $J<0$
the ferromagnetic phase of the effective spin-1 chain is expected
to be realized (for the original ladder it is the half-Ferro state with ground state total spin $S^T=L/2$).

So far, using analytical approaches, we have established ground-state phases of rung-alternated Heisenberg spin-1/2 ladder in the limiting cases of: $0<J \ll J_{\bot}$ and $J\gg J_{\bot}$, where
a Haldane phase of effective $S=1$ spins formed along the odd rungs and a
rung-singlet phases are stabilized respectively. The conjectured sequence of phases with changing $J$ is depicted in Fig. \ref{fig:modelest}. 

To check whether there are additional phases for intermediate values of $J\sim J_{\bot}$ or there is a direct phase transition between $H_1$ and RS states, we will use numerical approaches.

\section{Honeycomb-ladder model}

\begin{figure}%[ht]
\includegraphics[width=5.0cm]{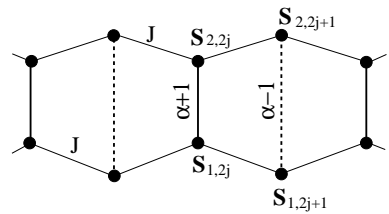}
\caption{  Geometry of {honeycomb-ladder} model. For $\alpha=0$ rung-alternated ladder presenting in Fig. 1 is recovered.}
\label{fig:2}
\end{figure}

In this section we will study a slightly generalized case of the rung-alternated ladder, introducing the additional parameter $\alpha$ that relaxes the condition of equal absolute values of exchanges along even and odd rungs. This way we obtain a $SU(2)$ symmetric version of the so called honeycomb-ladder model~\cite{Kara},
%%%%%%%%%%%%%
\begin{eqnarray}
\label{Themodel1}
H=J_{\bot} \sum^L_{i=j} [(-1)^j +\alpha ]{\bf S}_{1,j}{\bf S}_{2,j}+ J\sum^L_{j=1,l}  {\bf S}_{l,j}{\bf S}_{l,j+1}\,\,.
\end{eqnarray}
%%%%%%%%%%%%%
To simplify notations in the following we will set $J_{\bot}=1$.

For $\alpha \in (-1,1)$ the spin system is frustrated, because the
number of antiferromagnetic bonds per ladder plaquette is odd. We will study
the ground-state phases for arbitary values of $\alpha$ and $J$ (both
negative and positive). The simpler case to start from is $J<0$.

\subsection{Honeycomb-ladder with ferromagnetic legs}
For $J<0$ we can estimate {  the} boundary of {  the} ferromagnetic phase. We derive {  the} single-particle dispersions, composed of 4 branches, since {  the} unit cell contains 4 spins,
\begin{eqnarray}
\varepsilon_1(k)&=&\frac{\alpha}{2}+J\cos{k}, \,\,\,\, \varepsilon_2(k)=\frac{\alpha}{2}-J\cos{k}\nonumber\\
\varepsilon_3(k)&=&-\frac{\alpha}{2}-\frac{ \sqrt{ 2+J^2+J^2\cos{2k} } }{\sqrt{2}}\nonumber\\
\varepsilon_4(k)&=&-\frac{\alpha}{2}+\frac{ \sqrt{ 2+J^2+J^2\cos{2k} } }{\sqrt{2}}.
\end{eqnarray}

{  the} chemical potential is attached to the minimum of $\varepsilon_1(k)$ band, realized at $k=0$, $\mu=\alpha/2+J$.
By looking at the single magnon instability we estimate the boundary of the fully polarized state, equating {  the} chemical potential to the minimum of $\varepsilon_3(k)$ band, which is given for any system size by 
\begin{equation}
\alpha_{FM}=-\sqrt{1+J^2}-J \,\, \simeq -\frac{1}{2|J|} \,\, \mathrm{for}\,\,\,|J|\gg 1.
\end{equation}

Next, for $|J|\ll 1$ the single-particle dispersion bands become flat and we can estimate {  the} transition {  from the ferrimagnetic} state into the half-Ferro state with total $S=L/2$ by equating {  the} chemical potential with the maximum of $\varepsilon_3(k)$ band realized at $k=\pm\pi/2$,
\begin{equation}
\alpha_{c}=-J-1+O(J^2).
\end{equation}

Transitions induced by changing $\alpha<0$ can be understood in a simple
way for $|J|\ll 1$. In the half-Ferro state spins on even sites form rung-singlets, whereas on odd sites spins form
effective spins $S=1$. The effective spin-1 chain, formed by spins on odd sites, is in the fully polarized phase as follows for $J<0$ from Eq. \ref{Spin1}. The interaction between the spins belonging to even sites along the leg direction is mediated by intermediate $S=1$ spins (all of which are ponting in the same, spontaneously chosen, direction due to the fact that they are in fully polarized ground state) and this interaction is ferro since $J<0$. Hence, the system is equivalent to the direct product of the ferromagnetic state of the spin-1 chain (formed by spins belonging to odd sites) and a 2-leg spin-ladder with antiferromagnetic exhcange along rung and ferromagnetic exchange along legs\cite{Vekua03} (the ladder is formed by spins belonging to even sites). Decreasing $\alpha$ weakens the antiferromagnetic coupling along the rung and the effective magnetic field (produced by the fully polarized neighbouring spin-1 chain) induces two consequtive second order commensurate-incommensurate transitions, first one from rung-singlet state of the two-leg ladder (with ferromagnetic legs and antiferromagnetic rungs) to an intermediate Luttinger liquid state with finite polarization and then to fully polarized state\cite{Vekua04}.

\begin{figure}%[ht]
\includegraphics[width=4.0cm]{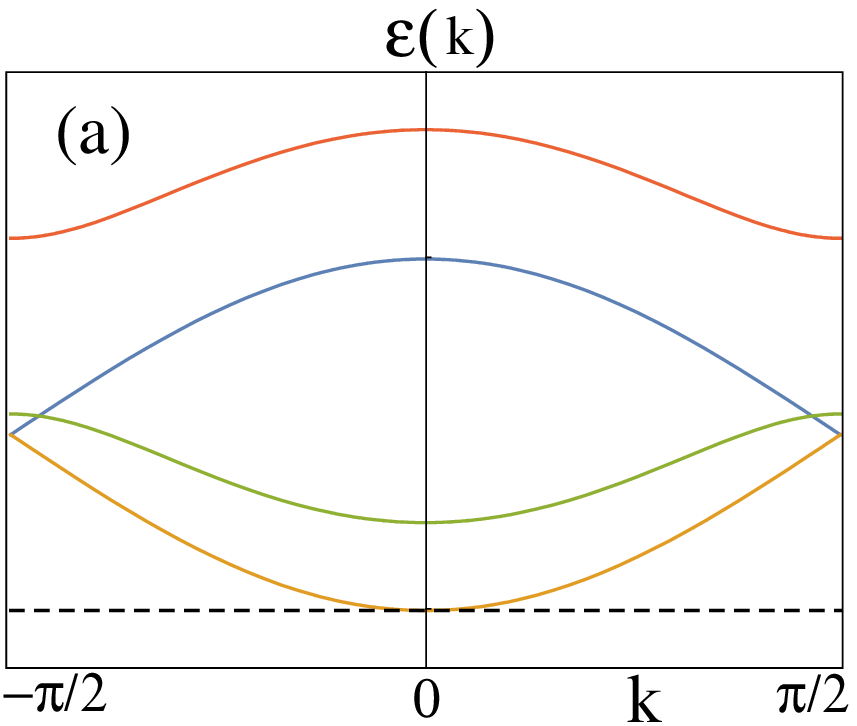}
\includegraphics[width=4.3cm]{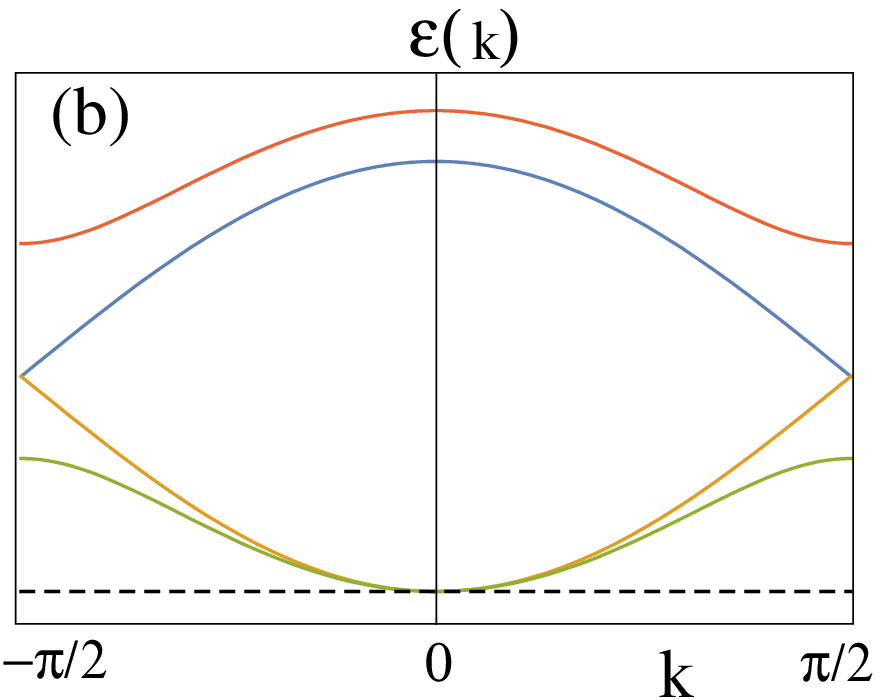}
\includegraphics[width=4.1cm]{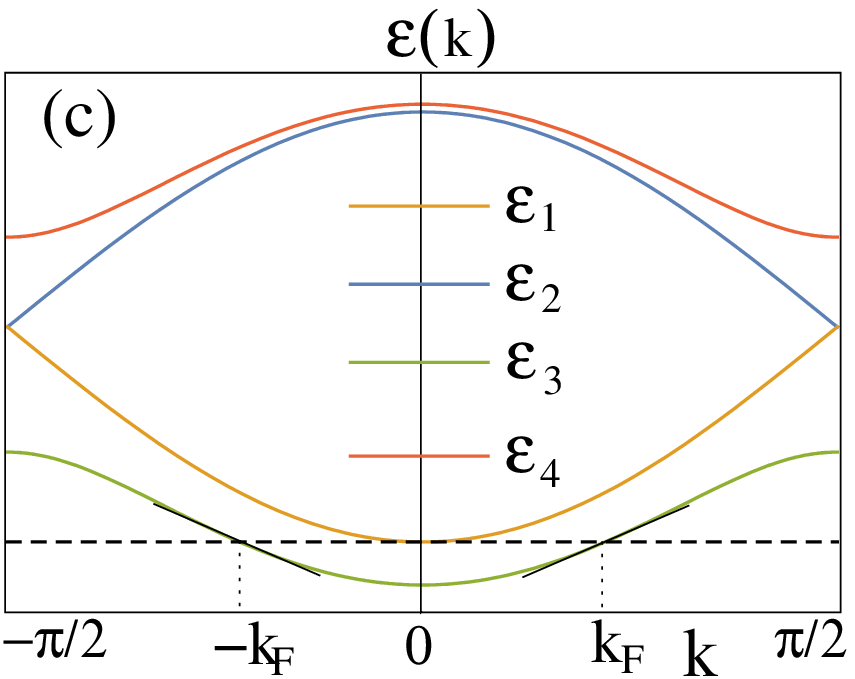}
\includegraphics[width=4.2cm]{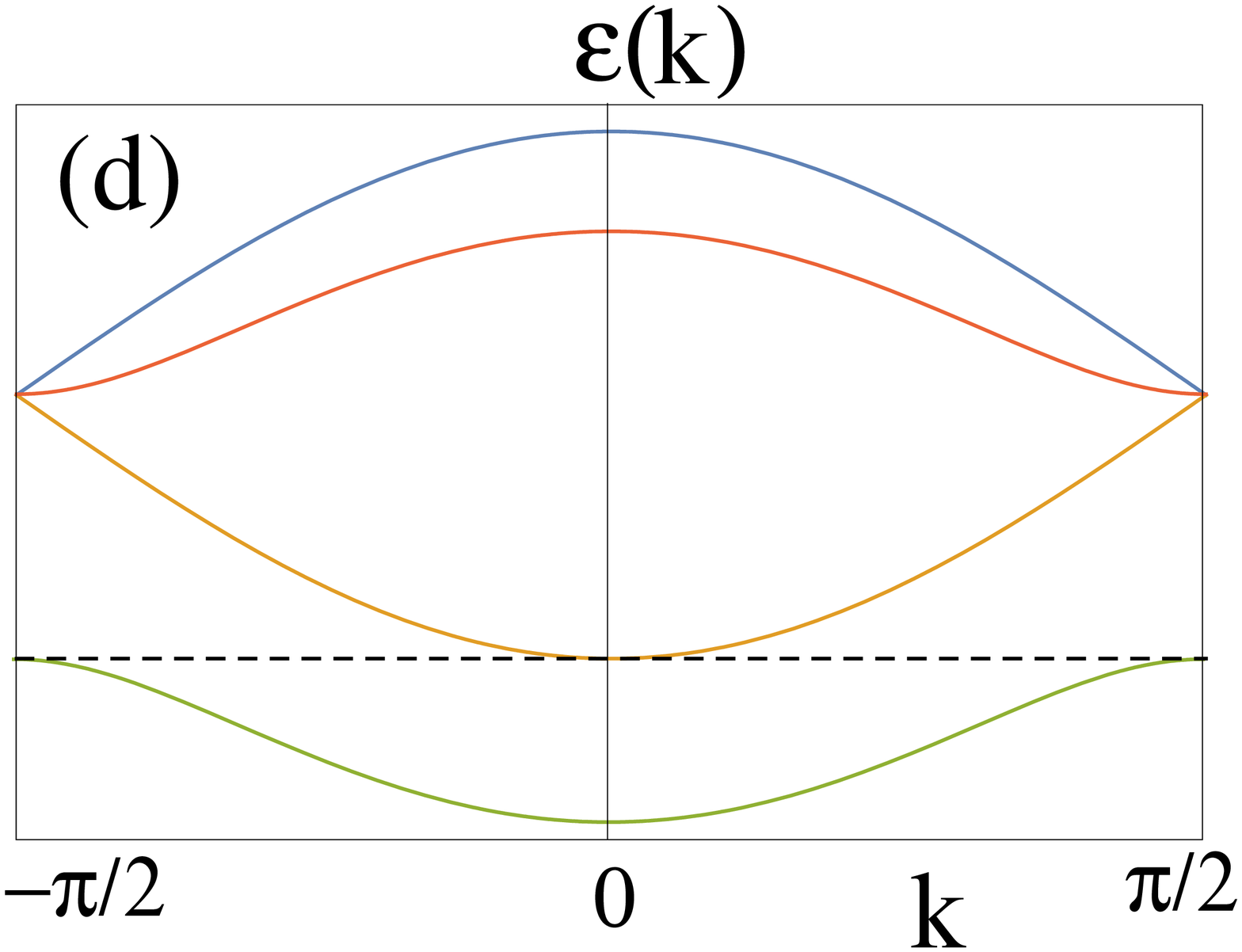}
\includegraphics[width=4.2cm]{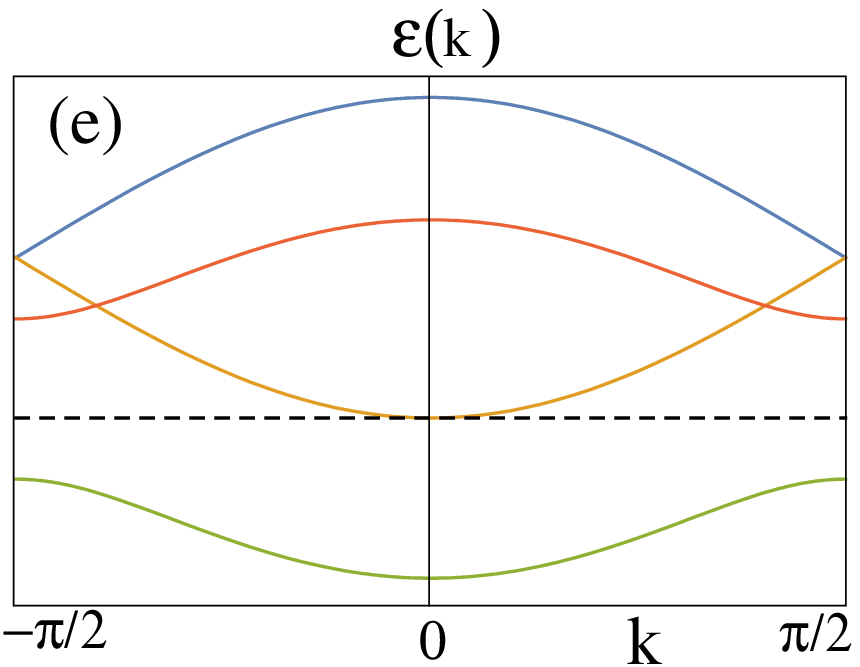}
\caption{ (Color online) Single particle dispersion relations (a) deep in ferro state, (b) at the boundary of ferro and ferri states, (c) inside ferri state, (d) at the boundary of ferri and half-ferro states {  and} (e) inside {  the} half-Ferro state. {  the} dashed line indicates {  the} chemical potential.}
\label{fig:dispersions}
\end{figure}
Single-magnon (equivalently hard-core boson) dispersions {  as} presented in
 Fig. \ref{fig:dispersions} shed light on the evolution of the low-energy
 excitation spectrum of model Eq.(\ref{Themodel1}) as function of $\alpha$ for
 $\alpha<0$ and $J<0$ and fully confirm the above picture anticipated
 from the interpenetrating spin-1 chain and {  the} two-leg ladder with
 ferromagnetic legs and antiferromagnetic rung exchanges. For
 $\alpha<-\sqrt{1+J^2}-J$ system is in {  the} ferromagnetic state and
 {  the} low energy excitation is {  a} conventional ferromagnetic magnon,
 with $\sim\cos{k}$-like gapless quadratic dispersion shown in
 Fig. \ref{fig:dispersions} (a). At the boundary between ferromagnetic and
 ferrimagnetic phases there are two quadratic {  dispersions}, two { 
 kinds} of magnons with the mass ratio of $\sqrt{2}$ presented in
 Fig. \ref{fig:dispersions} (b). Inside {  the} ferrimagnetic state the
 ground state total spin changes continuously $L/2\le S^T\le L$ and there are
 two different kinds of dispersions: one quadratic at low momenta and {  a}
 second one that is linear shown at $\pm k_F$ it Fig. \ref{fig:dispersions}
 (c), giving Luttinger liquid like properties. At the boundary between
 ferrimagnetic and half-Ferro states again two quadratic in momenta
 low energy dispersions are present shown in Fig. \ref{fig:dispersions} (d)
 and inside the half-Ferro state with  $S^T=L/2$ only one magnon
 branch with  $\sim\cos{k}$-like quadratic low energy dispersion remains 
 as shown in  Fig. \ref{fig:dispersions} (e). Both phase transitions, from
 ferromagnetic to ferrimagnetic and from ferrimagnetic to half-Ferro states are second order commensurate-incommensurate phase transitions, where {  the} linear mode disappears in favor of {  a} quadratic dispersion. The overall gapless quadratic mode remains in the background, due to {  the} spontaneously broken $SU(2)$ symmetry in all phases where {  the} ground state is not a global spin singlet.
 This makes $SU(2)$ symmetric honeycomb-ladder model Eq.(\ref{Themodel1}) for
 $J<0$ {  a} very attractive {  and} simple case to study the behavior
 known as unsaturated ferromagnetism, an effect that has been noticed to occur
 in other frustrated systems\cite{Sun,Giamarchi,Shimokawa,Nakano,Takano}. With
 further {  increase of} $\alpha$ to positive values there is a phase transition from half-Ferro to rung-singlet state, not captured by {  the} single particle picture.

\subsection{Effective model for $\alpha\simeq 1$ and $|J| \ll  1$}

\begin{figure}%[ht]
\includegraphics[width=8.0cm]{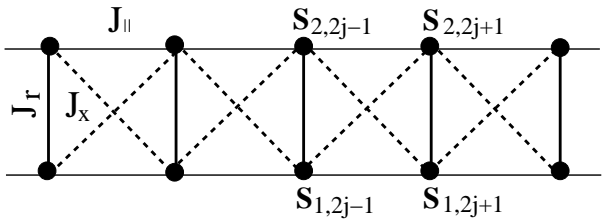}
\caption{ Effective model for spins belonging to odd rungs valid for $\alpha\simeq 1$ and $J\ll 1$, where spins belonging to even rungs form approximate rung-singlet states. Phase transitions are expected for $J_{r}<0$.  }
\label{fig:modeleffective}
\end{figure}

For $\alpha \simeq 1$ and $|J|\ll 1$ we can derive an effective model for spins on odd rungs by integrating out spins belonging to even rungs that are in rung-singlet states. To capture the difference between $J<0$ and $J>0$ cases we have to go beyond the lowest {  (second)} order in inter-rung coupling $J$ that we treat in perturbation theory.
To the third order in $J$ the effective ladder model formed by spins belonging to odd rungs is given by the the following hamiltonian (see Fig. \ref{fig:modeleffective}),

%%%%%%%%%%%%%
\begin{eqnarray}
\label{Themodeleffective}
H_{eff}=&&J_{||} \sum^{L/2}_{j=1} [{\bf S}_{1,2j-1}{\bf S}_{1,2j+1}+ {\bf S}_{2,2j-1}{\bf S}_{2,2j+1}]\nonumber\\
&+&J_{\mathrm x} \sum^{L/2}_{j=1} [ {\bf S}_{1,2j-1}{\bf S}_{2,2j+1}+  {\bf S}_{1,2j+1}{\bf S}_{2,2j-1} ] \nonumber\\ 
&+&J_{r}\sum^{L/2}_{j=1} {\bf S}_{1,2j+1}{\bf S}_{2,2j+1}
\end{eqnarray}
%%%%%%%%%%%%%

where
\begin{eqnarray}
\label{Themodeleffectiveparameters}
 J_{||}&=&-\frac{J^2}{2(\alpha+1)}- \frac{J^3}{4(\alpha+1)^2} \nonumber\\
J_{\mathrm x}&=&\frac{J^2}{2(\alpha+1)}+ \frac{3J^3}{4(\alpha+1)^2}\nonumber\\
J_{r}&=&    \frac{\alpha^2+J^2-1}{\alpha+1}+ \frac{3J^3}{2(\alpha+1)^2}.
\end{eqnarray}

Interchanging legs with diagonals of {  the} effective model presented in
Fig. \ref{fig:modeleffective} by interchanging spins on every other rung, the
model Eq.(\ref{Themodeleffective}) for parameters given in
Eq. (\ref{Themodeleffectiveparameters}) is equivalent to {  a} two-leg spin
ladder with antiferromagnetic legs, ferromagnetic diagonals and rung exchange
that changes from antiferromagnetic to ferromagnetic with decreasing
$\alpha$. For {  the} two-leg antiferromagnetic ladder weakly coupled by
competing diagonal and rung exchanges (where bosonization is applicable) three
phases are expected to be stabilized with decreasing $\alpha${  :}
rung-singlet, dimer and Haldane phase \cite{Balents}. Parameters of our
effective model are outside {  the} weak-coupling {  limit}, but later with the help of numerical simulations we will show that the same sequence of phases are also realized in our effective model Eq.(\ref{Themodeleffective}). In particular, the dimerization pattern of the original ladder model will be dimers formed along next nearest neighbor diagonals, involving spins belonging to odd rungs.

\subsection{Vicinity of $\alpha\simeq -1$, $0<J\ll 1$}

For $J>0$ and $0<J\ll 1$ we can as well estimate possible ground states around $\alpha \simeq -1$, where we can borrow the results from the mixed diamond chain \cite{Hida09,Hida14}.
For $J=0$ and $\alpha=-1$ the spins belonging to even rungs are disconnected and spins belonging to odd rungs in the ground state are in {  the} spin-triplet configuration, forming $S=1$ spins. 
For $0<J\ll 1$ there is a competition in the nature of the exchange between
the spins belonging to the even rungs: for $\alpha>-1$ the direct exchange
$S=1$  is antiferromagnetic, whereas the exchange mediated by nearby $S=1$
spins is ferromagnetic. For $\alpha<-1$ there is no such competition {  and effective spins $S=1$ are formed on each rung}. 

\begin{figure}%[ht]
\includegraphics[width=8.0cm]{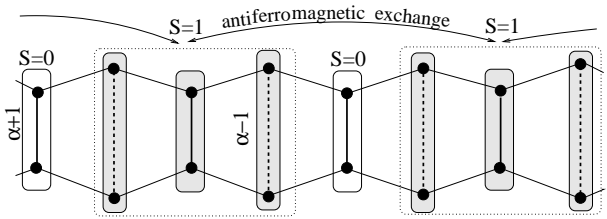}
\caption{Cartoon of one of the possible ground state configurations of {  the} Haldane-Dimer phase that can be realized for $\alpha\simeq -1$, $0<J\ll 1$. Spins encircled by open rectangles form approximate rung-singlet state, whereas those encircled by shaded rectangles form rung-triplet states. 6 spins encircled by {  the} dotted rectangle form effective $S=1$ spins which are connected via intermediate singlets to produce {  an} effective Haldane chain. In {  the} mixed-diamond chain the singlets depicted above are exact eigenstates and they do not mediate any exchange among {  the} effective $S=1$ spins.}
\label{fig:Hida}
\end{figure}

One possibility that the above mentioned competition for $\alpha>-1$ gets
resolved is that some of the even rungs choose to be in triplet state and
others in singlet state periodically alternating as happens in {  the}
mixed-diamond chain \cite{Hida09} where {  consecutive} odd number of $M$
rung-triplets (coupled antiferromagnetically with each other by $J>0$) will
be {  sandwiched} between the rung-singlets. 
Coupling {  an} odd number $M$ of $S=1$ spins by antiferromagnetic exchange
and assuming open boundary conditions, the $M$-rung segment will be in the
triplet state in the ground state, forming an effective $S=1$ spin. The
approximate rung singlets in the case of rung-alternated ladder (as opposed to
the exact rung singlets realized in mixed-diamond chain  \cite{Hida09} that
cut the chain) will mediate {  an} effective antiferromagnetic exchange
among the above mentioned effective $S=1$ spins formed by $M$-rung segments, giving rise to generalized Haldane states with {  an} enlarged unit cell composed of $M+1$ ladder plaquettes. Such {  a} state for $M=3$ is depicted in Fig. \ref{fig:Hida} and called Haldane-dimer.  

There are in total 5 different ground states in {  the} mixed-diamond chain when changing {  the} equaivalent of $\alpha$ from $\alpha>-1$ to $\alpha<-1$ with $M=1,3,5,7$ and $M=\infty$. {  The} $M=1$ case is equivalent to {  the} H$_1$ state, a Haldane state of $S=1$ spins formed on odd rungs, and {  the} $M=\infty$ case is equivalent to the Haldane state of the effective $S=1$ spins formed on every rung. For the spin-1/2 ladder with alternated rungs $M=1,3,5$ and $M=\infty$ are suggested to have finite extent in the presence of exchange anisotropy \cite{Hida14}.

\section{Numerical results}

\subsection{Rung-alternated ladder}

Initially we will present our numerical data for $\alpha=0$ case corresponding
to {  the} rung-alternated ladder model Eq. (1). We use both Lanczos simulation and  the density matrix renormalization group (DMRG) approach
\cite{White92,Scholl05} in order to access large system sizes.
For the case of ferromagnetic legs we systematically obtain (both using large scale DMRG as
well as Lanczos {  algorithm} for both periodic and open boundary
conditions) that {  the} ground state belongs to the multiplet with total
{  spin half} of the maximal possible value, $S^T=L/2$ for any $J<0$.

\begin{figure}%[ht]
\includegraphics[width=8.0cm]{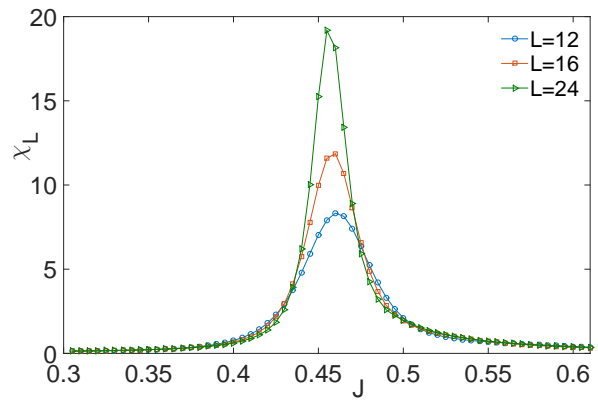}
\caption{(Color online) Fidelity susceptibility of {  the} rung-alternated ladder with antiferromagnetic legs for periodic boundary conditions and 3 different system sizes. In DMRG calculations periodic boundary conditions are assumed (restricting considerably {  the} available ladder lengths) to avoid degeneracies of the Haldane-like states due to edge spins for the open boundaries.}
\label{fig:FSa0}
\end{figure}

In the following we will discuss $J>0$, where we predicted at least two different phases in the limiting cases $J\ll 1$ and $J\gg 1$ respectively. {  In Fig.} \ref{fig:FSa0} we plot {  the} fidelity susceptibility\cite{VenutiZanardi07,You+07,Gu10rev,us} with changing control parameter $J$ for different system sizes. 

\begin{equation}
\label{FSE}
\chi_L=-\frac{2}{L}\lim_{\delta J\to 0} \frac{\ln {| \langle \psi_0(J)|  \psi_0(J+\delta J) \rangle |}}{(\delta J)^2},
\end{equation}
where $|\psi_0(J)\rangle$ is the (non-degenerate) ground state wavefunction for the corresponding parameter $J$.
We see that there is a well pronounced peak in {  the} fidelity susceptibility and {  the} height of the peak increases with system size, whereas the width decreases. We extrapolate the location ot the peak to $J=J_{c1} \simeq 0.45$ in the thermodynamic limit. Thus, we can estimate the extent of the H$_1$ phase for $\alpha=0$ as $0<J\le J_{c1}$. Note, the rather similar estimate of $J_{c1}$ follows from the position of the level-crossing of the lowest excited states, which are triplet with momentum $k=\pi$ in H$_1$ phase ($J<J_{c1}$) and triplet with $k=0$ in its neighboring phase ($J>J_{c1}$).

\subsection{Honeycomb-ladder with ferromagnetic legs: $J<0$}

\begin{figure}%[ht]
\includegraphics[width=9.0cm]{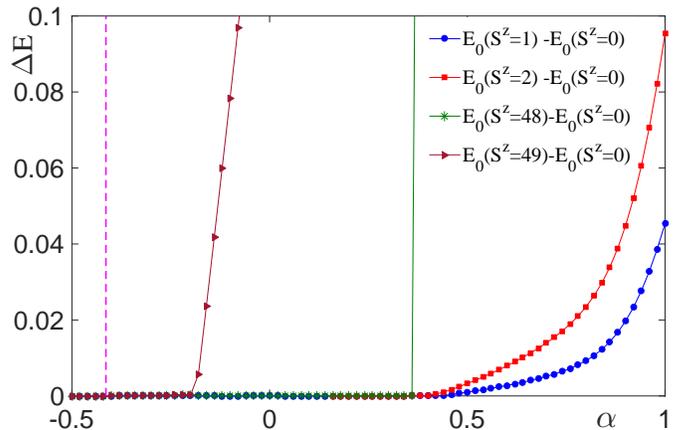}
\caption{(Color online) DMRG results for the lowest energy levels in different  total $S^z$ subspaces relative to the lowest energy in the $S^z=0$ subspace as function of $\alpha$ for $J=-1$ and $L=96$ rungs for open  boundary conditions. The dashed line indicates the boundary of the ferromagnetic state.}
\label{fig:FerroGaps}
\end{figure}

We start by presenting our numerical data with the case of ferromagnetic
legs. In Fig. \ref{fig:FerroGaps} we plot for {the} honeycomb-ladder with ferromagnetic legs the ground state multiplicity as function of $\alpha$ for 
$J=-1$, which is a typical behavior in the whole $J<0$ region. The boundary
of the ferromagnetic phase is captured exactly from the spin-wave
instability (indicated by dashed vertical line in Fig. \ref{fig:FerroGaps}). Here we only present the data from which we determine the boundaries of the half-Ferro state.

 We see from this plot that for $J=-1$ {  the} half-Ferro state $S^T=L/2$ is sandwiched between $\alpha\simeq -0.2$ and $\alpha \simeq 0.36$. Note, for $0.36<\alpha<0.4$ the excitations from the singlet ground state to low total spin states e.g. to states with $S^T=1$ and $2$ become practically gapless, thus we cannot rule out the existence of an intermediate thin phase between half-Ferro and rung-singlet states based on our numerical data.  

For values of $\alpha<-0.2$ energies of the lowest states with $S^z>L/2$
merge gradually with the ground state (only one state $S^z=L/2+1$ is
indicated in Fig. \ref{fig:FerroGaps}) until the energy of the fully polarized state with $S^z=L$ becomes degenerate with the ground state energy for $\alpha\le \alpha_{FM}$, $\alpha_{FM}(J=-1)= 1-\sqrt{2}\simeq-0.414$.

\subsection{Honeycomb-ladder with antiferromagnetic legs: $J>0$}

\begin{figure}%[ht]
\includegraphics[width=8.0cm]{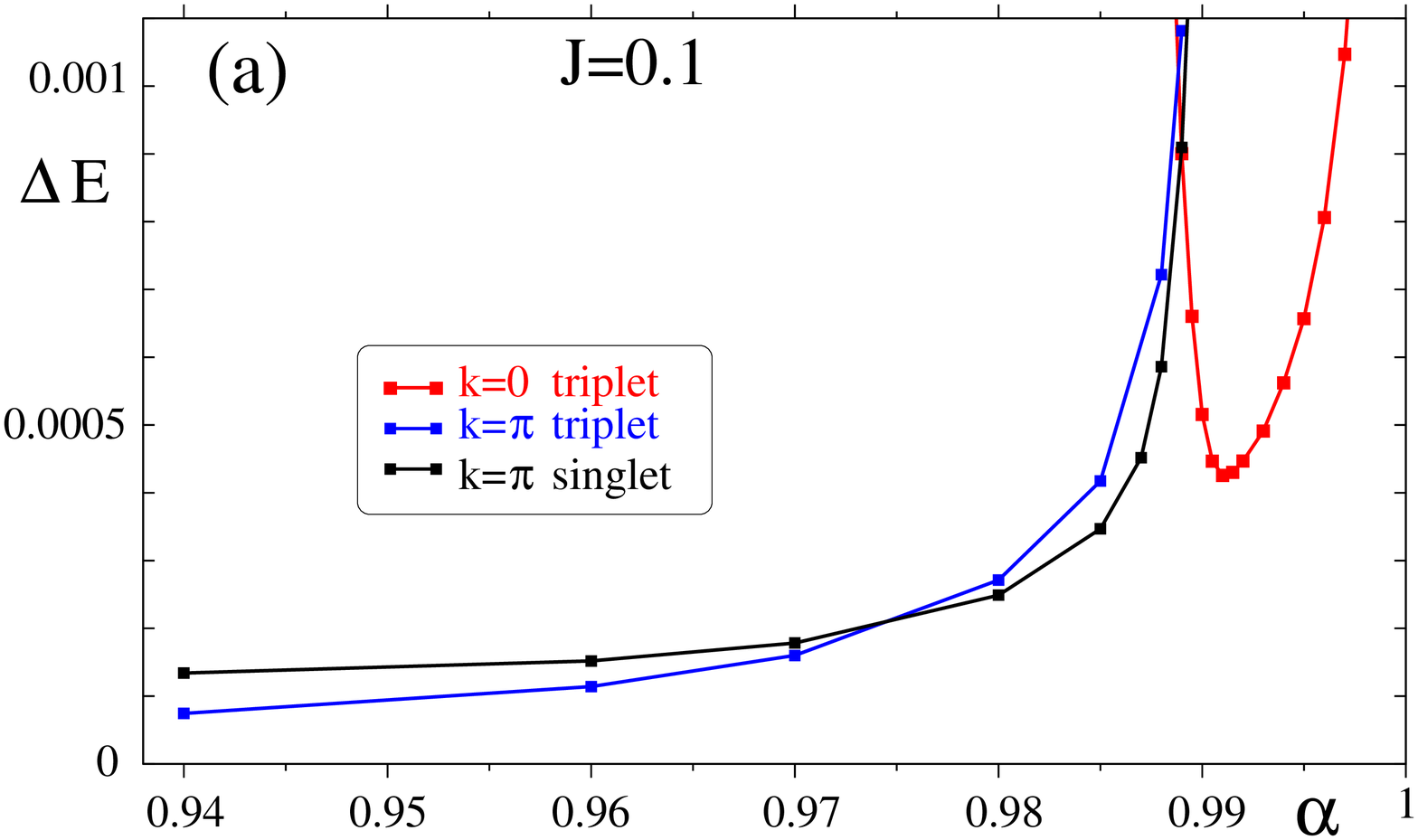}
\includegraphics[width=8.0cm]{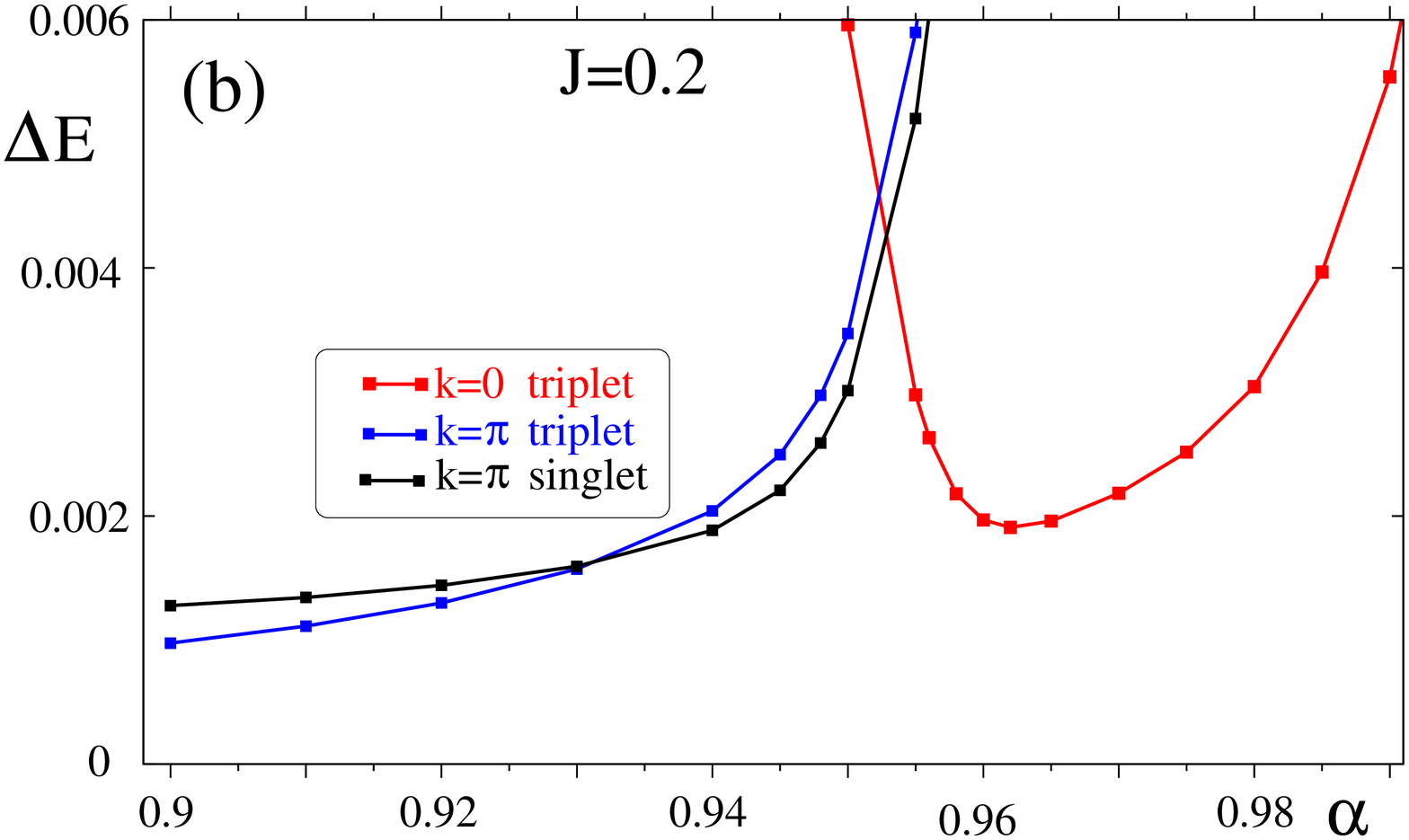}
\caption{ (Color online) Lanczos {  results} of the lowest excited states
  relative to the ground state obtained for $L=12$ rungs of the effective
  model Eq.(\ref{Themodeleffective}) for (a) $J=0.1$ and (b) J=0.2. {  A} similar picture is expected to hold for {  the} Honeycomb-ladder model with $L=24$ rungs for the same extent of $\alpha$ and $J$. Periodic boundary conditions are used that allows to assign a definite lattice momentum to each level.}
\label{fig:RSDH}
\end{figure}

For $J>0$ case we start presenting numerical data near the point $\alpha\simeq 1$ for small $J$ with changing $\alpha$. To distinguish different phases it is usefull to start from looking at the gap between the ground state and lowest excited states. In Fig. \ref{fig:RSDH} we depict the lowest excited states as function of $\alpha$. We use the effective model Eq. (\ref{Themodeleffective}) to reach system sizes of $L=12$ rungs, that is equivalent to $L=24$ rungs for the Honeycomb-ladder model Eq. (\ref{Themodel1}). We have checked that for available system size (up to $L=12$ for Honeycomb-ladder model) agreement between the low energy levels of effective and full models is perfect for small $J$ values. In Fig.\ref{fig:RSDH} we present the level spectroscopy results for $J=0.1$ (a) and $J=0.2$ (b) for the effective model Eq. (\ref{Themodel1}).
In an antiferromagnetic ladder with a uniform antiferromagnetic exchange the lowest excited state in the rung-singlet phase is a triplet state with wavevector $k=\pi$ in units of the ladder lattice constant. Since in our model the unit cell is made of two plaquettes, in the rung-singlet phase of Honeycomb-ladder (in the phase that is adiabatically connected with rung-singlet phase of the uniform ladder, but with a unit cell half of the Honeycomb-ladder model) the lowest excited triplet should have momentum $k=2\pi$ in the units of the Honeycomb-ladder unit cell that is equivalent to $k=0$ momentum.

We see that with decreasing $\alpha$ below $\alpha=1$ the gap to the lowest excitation (triplet state with $k=0$ momentum) shows a minimum and then with reducing $\alpha$ this lowest triplet excitation level crosses with the lowest excited singlet state that has momentum $k=\pi$. Note, for any $J>0$ and any $\alpha$ the ground state is a spin singlet state with $k=0$ momentum. There is a finite extent in $\alpha$ where the lowest excited state is a singlet state with $k=\pi$. With further reducing $\alpha$ there is a level-crossing between the lowest spin singlet excitation with $k=\pi$ and spin triplet excitation with $k=\pi$. The spin triplet excitation with $k=\pi$ in units of Honeycomb-ladder unit cell is the lowest excitation on top of the Haldane state that is defined on the effective spin-1 chain with the same unit cell as the original microscopic model. One can use the abovementioned two level crossings in excited states to estimate the stability region of the intermediate dimer phase. In fact with increasing system size the singlet excitation at $k=\pi$ should get degenerate with the ground state singlet in the dimerized phase. The boundary between the rung-singlet and dimer states can be estimated from the position of the minimum of the gap of the $k=0$ triplet state.

Comparing the energy levels of the effective ladder model Eq.(\ref{Themodeleffective}) for different system sizes with $L\le 12$ rungs we see that the energy of singlet state with $k=\pi$ momentum decreases faster, with increasing the system size, than energies of the triplet states in the parameter region where the dimerized phase is expected.

\begin{figure}%[ht]
\includegraphics[width=8.5cm]{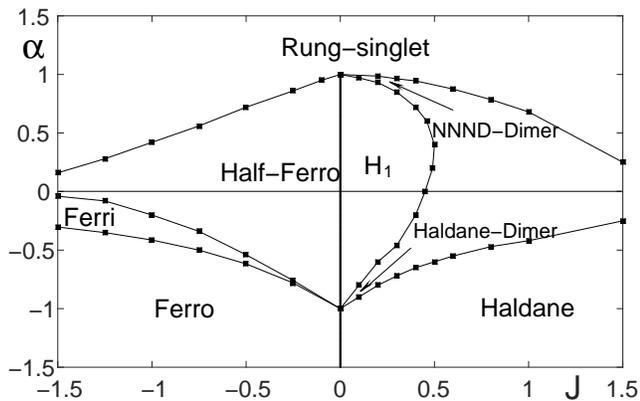}
\caption{ Numerical ground-state phase diagram of the honeycomb-ladder model in the parameter plane $(J,\alpha)$. NNND-Dimer stands for next nearest neighbor diagonal dimer phase where dimers are formed along next nearest neighbor diagonals involving spins of odd rungs. In the vicinity of $\alpha=-1$ and for $0<J\ll 1$ before {  the} transition from Haldane-Dimer to Haldane phase additional phases may occur (e.g. with $M=5$ and $M=7$ as discussed in previous sections). In dimer phases ground states are doubly degenerate in the thermodynamic} limit.
\label{fig:PD}
\end{figure}

The numerical ground-state phase diagram of {  the} honeycomb-ladder model obtained with the help of DMRG simulations is presented in Fig. \ref{fig:PD}.
For $J<0$ there are four different phases realized with decreasing $\alpha$: rung-singlet, half-ferro, ferrimagnetic and ferromagnetic. For the case $J>0$ {  the} rung-singlet state, {  the} H$_1$ and {  the} conventional Haldane state and different dimerized states: NNND and Haldane-Dimer are realized. Between the NNND-Dimer and Haldane-Dimer states we can not locate numerically the phase transition line, neither can we exclude emergence of {  an} intermediate (gapless) state located around $\alpha=0$. 

It is worth noting that the topology of the H$_1$ phase realized for $J\ge 0$ can be captured by studying one plaquette of the ladder, $L=2$. Consider e.g. {  the} $\alpha=0$ case. For this case for $J<1/\sqrt{2}$ a triplet state is realized as ground state, whereas for $J>1/\sqrt{2}$ the ground state becomes {  a} singlet, a direct product of the singlet states on {  the} first chain (2-site chain) and on the second chain (that is an exact {  eigenstate} for any $J$ in the case of a single plaquette). Hence at $J=1/\sqrt{2}$ there is a triplet-singlet level crossing in the ground state of one plaquette. The {  (threefold)} degeneracy of the ground state for small values of $J$ is {  a} particular case and {  omit:it} stems from the fact that there is only one effective spin 1 (formed on one of the two rungs). As soon as {  the} number of ladder plaquettes is increased and more than one effective spin 1 {  is}  formed on odd rungs the ground state becomes {  a} singlet (for periodic boundary conditions) for the whole range of $J>0$ and there is no level crossing in the ground state any more. However, when one {  assembles} many plaquettes into {  the} ladder geometry, instead of the level-crossing, one can identify the avoided level crossing in the lowest energy singlet states of the finite ladder (for system sizes $L\le 12$ rungs), that is located at $J\sim 0.5 $ (data not shown).

\begin{figure}%[ht]
\includegraphics[width=8.3cm]{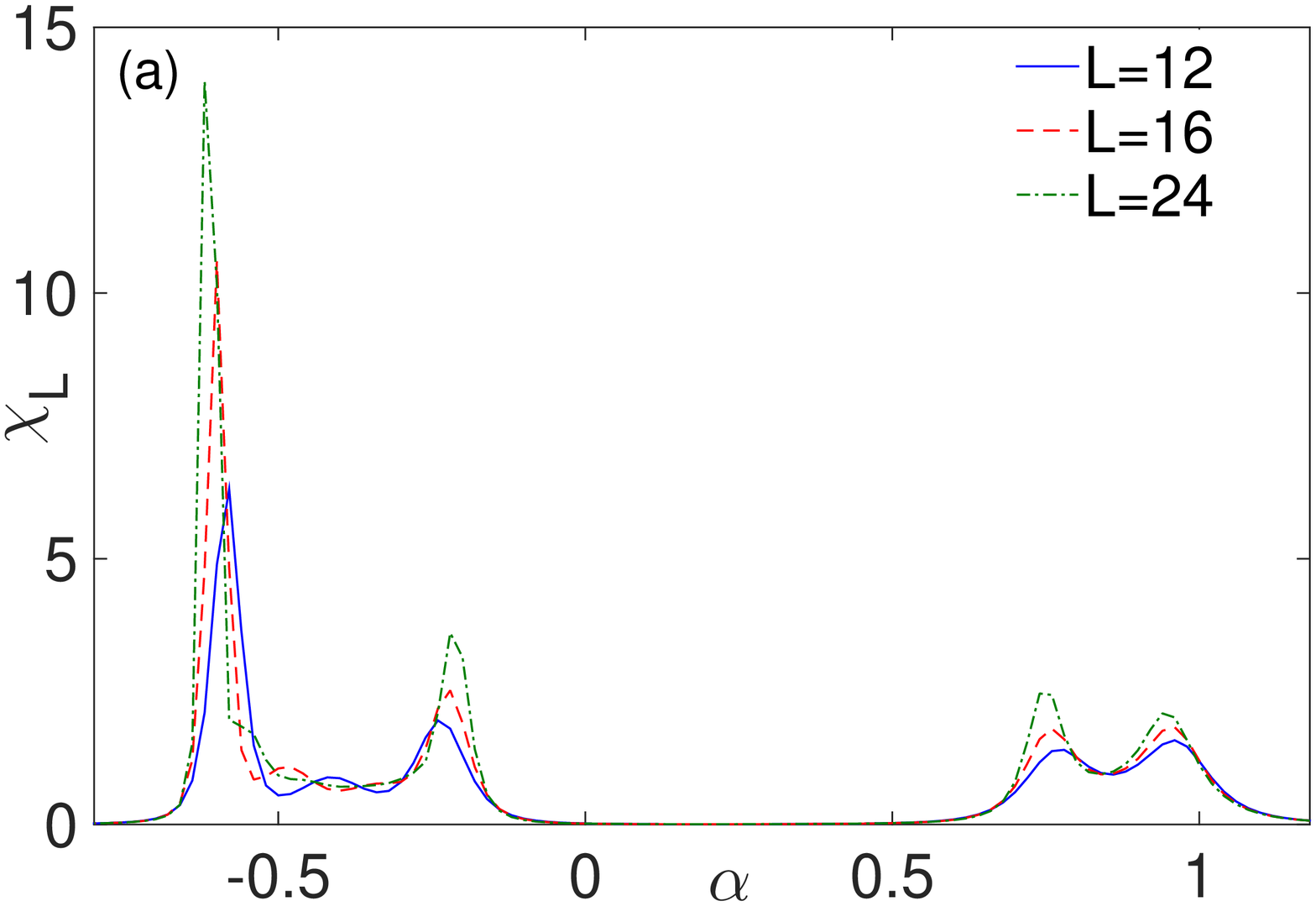}
\includegraphics[width=8.5cm]{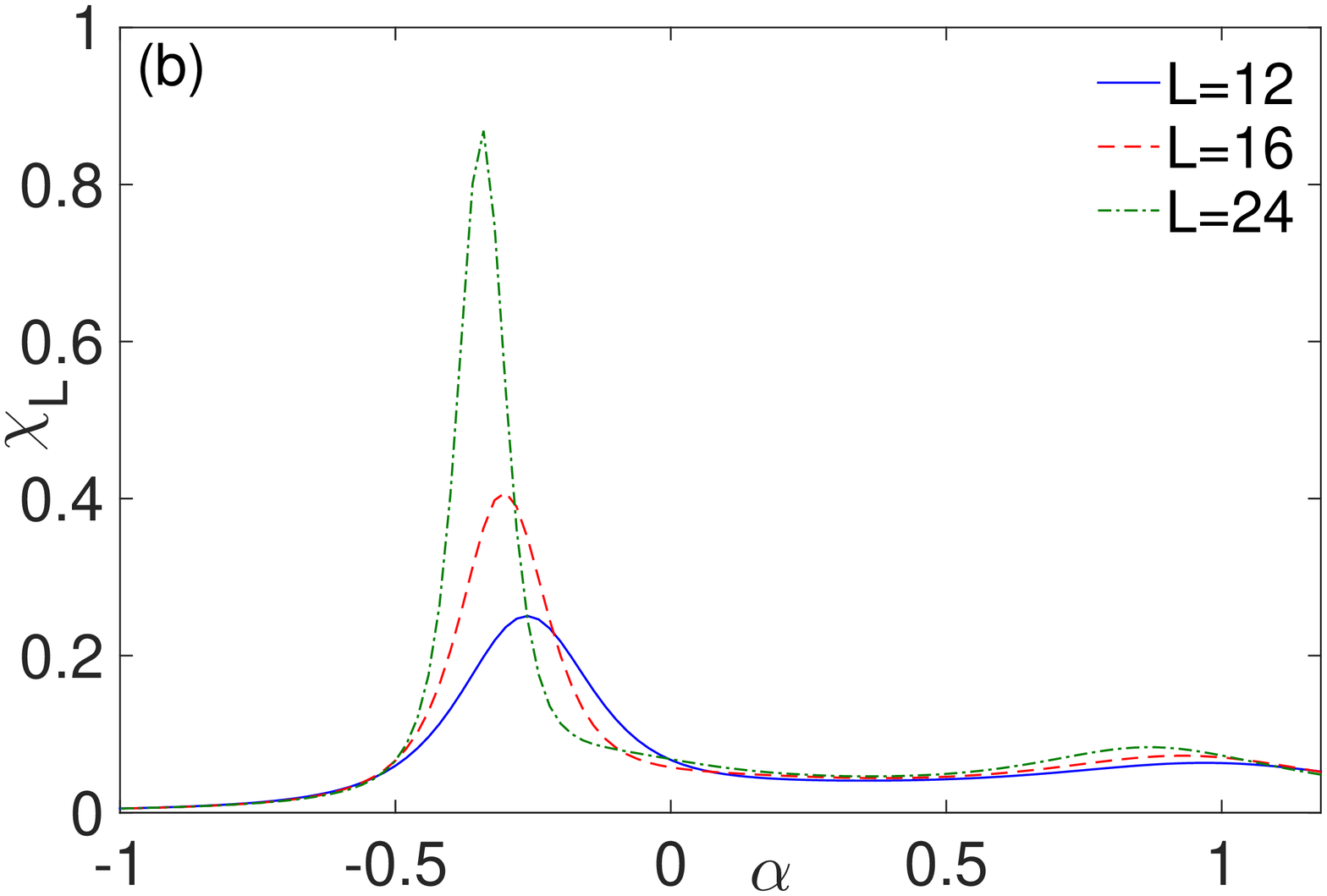}
\caption{(Color online) Ground state fidelity susceptibility per site as function of $\alpha$ for (a) $J=0.4$ and for $J=1$ showing two peaks. In DMRG calculations periodic boundary conditions are assumed to avoid degeneracies of the Haldane-like states due to edge spins for open boundaries.}
\label{fig:FS}
\end{figure}

The phase transition points indicated in Fig. \ref{fig:PD} for $J>0$ were obtained by studying the behavior of the fidelity susceptibility as function of $\alpha$ for different values of $J$ as presented in Fig. \ref{fig:FS}. For small values of $J$ (roughly $J<0.5$) {  the} fidelity susceptibility shows typically well pronounced four peaks, whereas for $J>0.5$ only two peaks are visible, one for positive and a second one  for negative $\alpha$. The peak for the $\alpha>0$ side becomes less and less pronounced with increasing $J>1$. 

\begin{figure}%[ht]
\includegraphics[width=9.0cm]{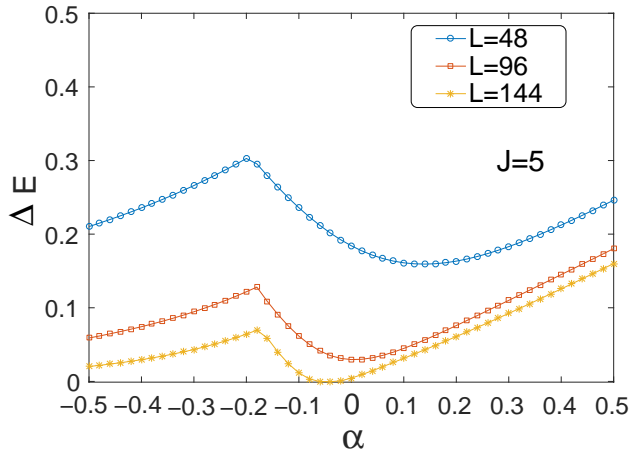}
\caption{Gap between the ground state and the first excited (triplet) state as function of $\alpha$ for $J=5$ obtained by DMRG using open boundary conditions.}
\label{fig:gapJ5}
\end{figure}

To describe the regime corresponding to $J\gg 1$ in Fig.\ref{fig:gapJ5} we present the behavior of the lowest excitation gap as function of $\alpha$ using DMRG for large value of $J=5$. In order to access large system sizes we use open boundary conditions. One can see that with decreasing $\alpha$ first there is a minimum in gap and then there is a cusp-like behavior. Using the finite system size data for systems with $L=48, 96$ and $144$ rungs the position of the gap minimum in the thermodynamic limit extrapolates clearly to negative values of $\alpha$. Starting from the rung-singlet phase, the gap decreases linearly with decreasing $\alpha$ and the position of the minimum of the gap we interpret as a boundary of the rung-singlet phase.

On the other hand, for $J\gg 1$, extending the bosonization analyses to $\alpha\neq 0$ gives that for $\alpha>0$ the rung-singlet phase smoothly evolves into the rung-singlet phase of the uniform antiferromagnetic ladder realized for $\alpha \gg 1$. For $\alpha< 0$ interestingly bosonization suggests the sequence of two consecutive second order phase transitions, first from rung-singlet to an intermediate dimer phase and then from dimer to Haldane phase with decreasing $\alpha$. Hence we expect to see two values of $\alpha<0$ where gap should close in the thermodynamic limit. Instead we see only one minimum in the finite-size gap data presented in Fig. \ref{fig:gapJ5}. The reason why we do not see the second minimum may be the fact that finite-size effects are still large (even for $L=144$ rungs). In addition, since we use open boundary conditions, it is difficult to separate true bulk gap from the boundary gap of the Haldane phase (realized on the left side from the kink in Fig. \ref{fig:gapJ5}). It is desirable to study the gap for periodic boundary conditions, however DMRG calculations become less accurate and only much smaller system sizes can be addressed.

\section{Conclusions}

We have studied the ground-state phase diagram of the rung-alternated $SU(2)$ symmetric spin$-1/2$ ladder. Both cases with ferromagnetic as well as antiferromagnetic leg exchanges have been considered. For the case of ferromagnetic legs we showed that a unique ferrimagnetic ground state emerges, with ground state magnetization equal to half of the maximum possible value, for arbitrary strength of the leg exchanges. The case of antiferromagnetic leg exchange is much richer and depending on the ratio of leg to rung couplings several different ground states can emerge starting from the Haldane phase H$_1$ for small leg couplings and ending with the rung-singlet phase for strong leg couplings. Based on Fig. \ref{fig:PD} it is tempting to speculate that dimer order extends to $\alpha=0$ and hence the intermediate phase of rung-alternated ladder can be dimerized, even though we have not succeeded in either directly measuring dimerization order, or finding a second singlet state as the lowest excited state of the finite chain or even resolving a finite excitation gap numerically.

We have as well studied a generalization of rung-alternated ladder: the spin$-1/2$ Heisenberg system on honeycomb-ladder lattice. For the case of ferromagnetic legs we have identified a peculiar Luttinger liquid ferrimagnetic state, where the ground state magnetization changes continuously as function of system parameters and low energy gapless excitations consist of two branches one of which is linear and another quadratic in momentum. For the case of antiferromagnetic leg couplings different short-range ground states, including those with possible Haldane-like topological order have been suggested to occur.

 This work has been supported by DFG Research Training Group (Graduiertenkolleg) 1729 and center for quantum engeneering and space-time research (QUEST). Work of F. A. was done while visiting Institute of Theoretical Physics, Leibniz University of Hanover. 
F. A. acknowledges grant from the ministry of science and technology of Iran and support from deputy of research and technology of university of Guilan. We thank S. Greschner for numerical assistance.

%%%%%%%%%%%%%%

\end{document}